\documentclass[12pt,a4paper,oneside]{article}
\usepackage[utf8]{inputenc}
\usepackage{amsmath}
\usepackage{bm}
\usepackage{amsfonts}
\usepackage{graphicx}
\usepackage{epstopdf}
\usepackage{hyperref}
\usepackage{array}
\usepackage{soul}
\usepackage{xcolor}
\usepackage{comment}
\usepackage{float}
\enlargethispage{\baselineskip}

\hypersetup{
     colorlinks   = true,
     citecolor    = blue
}

\newcommand*{\lh}[1]{\lambda_{h}}
 
\newcommand*{\lc}[1]{\lambda_{c}}
\newcommand*{\lo}[1]{\lambda_{0}}

\usepackage{jheppub_1}
\usepackage[english]{babel}

\usepackage{amsmath,amsfonts,amssymb}
\usepackage{graphicx}
\usepackage{subfigure}
\usepackage{siunitx}
\usepackage{xspace}
\usepackage{braket}
\usepackage{mathtools}
\usepackage{hyperref}
\usepackage{comment}
\usepackage[normalem]{ulem}
\usepackage[capitalize]{cleveref}
\hypersetup{
    colorlinks=true,
    linkcolor=blue,
    citecolor=blue,
    urlcolor=blue
}

\binoppenalty=\maxdimen
\relpenalty=\maxdimen 

\graphicspath{{./graphics/}}

\def\bea{\begin{eqnarray}}
\def\eea{\end{eqnarray}}
\def\l{\left}
\def\r{\right}
\def\d{\partial}

\providecommand{\abs}[1]{\lvert#1\rvert}
\providecommand{\bd}[1]{\boldsymbol{#1}}
\providecommand{\ro}[1]{\mathrm{#1}}

\DeclareSIUnit\parsec{pc}
\crefname{section}{section}{sections}
\Crefname{section}{Section}{Sections}
\crefname{appendix}{appendix}{appendices}
\Crefname{appendix}{Appendix}{Appendices}
\crefname{table}{Tab.}{Tabs.}
\Crefname{table}{Table}{Tables}

\makeatletter
\gdef\@fpheader{\strut}
\makeatother

\title{Bubble Nucleation from Boson Star Collapse
}
\preprint{SISSA 08/2025/FISI}

\author{Aleksandr Azatov,}
\author{Takeshi Kobayashi,}
\author{and Nicklas~Ramberg}

\affiliation[]{SISSA, International School for Advanced Studies, Via Bonomea 265, 34136 Trieste, Italy}
\affiliation[]{INFN Sezione di Trieste, Via Bonomea 265, 34136 Trieste, Italy}
\affiliation[]{IFPU, Institute for Fundamental Physics of the Universe, Via Beirut 2, 34014 Trieste, Italy
}

\emailAdd{aleksandr.azatov@sissa.it}
\emailAdd{takeshi.kobayashi@sissa.it}
\emailAdd{nramberg@sissa.it}

\abstract{We present a new classical mechanism for nucleation of bubbles of true vacuum. The mechanism arises when dense boson stars form in the false vacuum. As the boson stars collapse due to attractive self-interactions, the field inside the star cores is enhanced beyond the potential barrier. Subsequently the stars explode as true vacuum bubbles, and induce a cosmological phase transition. The mechanism raises the possibility that a vacuum that is stable against quantum tunneling can be vulnerable to ``astrophysical'' processes.}
\begin{document}

\maketitle\flushbottom

\section{Introduction}

Recent advances in gravitational wave astronomy \cite{Barausse:2020rsu,NANOGrav:2023hvm}
have sparked interest in the physics of cosmological first-order phase transitions \cite{Coleman:1977py,Linde:1980tt}, potentially leading to stochastic gravitational wave signals that are observable in current and future experiments.
These experiments can reach frequencies up to a few kHz, 
which correspond to transition scales much higher than the physics scales reachable by collider experiments.\footnote{A present-day frequency of kHz is tied to a transition scale of $ \sim 10^{10}\, \ro{GeV}$
in conventional scenarios. 
This relation however is modified in the mechanism we will present.}
The signal details depend strongly on the physics behind the phase transition, which usually proceeds via quantum 
tunnelling~\cite{Coleman:1977py} or thermal fluctuations~\cite{Linde:1980tt,LINDE1980289} creating supercritical bubbles which later expand and collide. In this paper, we 
would like to present a new classical mechanism for the creation of supercritical bubbles from the collapse of a boson star.

Boson stars \cite{Kaup:1968zz,Ruffini:1969qy} are clumps of Bose-Einstein condensate that are held in equilibrium by gravity or attractive self-interactions. 
They are known to arise in various early universe scenarios~\cite{Colpi:1986ye,Liddle:1992fmk,Kolb:1993zz,Khlebnikov:1999pt,Sikivie:2009qn,Chavanis:2011zi,Chavanis:2011zm,Chavanis:2025qcg,Chavanis:2017loo,Erken:2011vv,Davidson:2013aba,Davidson:2014hfa,Guth:2014hsa,Eby:2015hsq,Levkov:2016rkk,Schiappacasse:2017ham,Levkov:2018kau,Eggemeier:2019jsu,Chen:2021oot,Chan:2022bkz,Dmitriev:2023ipv,Gorghetto:2024vnp,Ralegankar:2024zjd}, and their physics has been extensively discussed especially in the context of axions
(see e.g.~\cite{Marsh:2015xka,Zhang:2018slz} for reviews).
One of the important consequences of boson stars is that their evolutionary track can end with a collapse, which drives the energy density in the core of the star to very large values. 
This is indeed the case for axions, whose attractive self-interaction renders axion stars unstable against collapse; 
the axion star collapse has been shown~\cite{Levkov:2016rkk} to be accompanied by emission of relativistic axions from the high-density core.

In this work, we study the collapse of boson stars consisting of a scalar field whose potential has non-degenerate minima. 
Considering the universe to be initially trapped in a false vacuum, 
we explore the possibility that a boson star collapse induces large 
energy densities, i.e., field excursions, to overcome the potential barrier inside the star's core.
We demonstrate that this process can actually create a supercritical bubble, which subsequently turns to expand and triggers a cosmological phase transition. To delineate the region of parameter space in which this mechanism 
operates, we perform numerical simulations of the classical 
field evolution on a lattice. These simulations explicitly 
demonstrate the formation of 
supercritical bubbles, and are complemented by qualitative arguments that support and clarify our numerical results.
We thus show that a boson star can classically seed a vacuum transition, even if the false vacuum is stable against quantum tunneling. 

Generally speaking, the process we investigate can be viewed as a true vacuum bubble creation driven by  
field–interaction dynamics. In this context, we would like to mention Refs. \cite{Easther:2009ft,Giblin:2010bd}, where a different type of field dynamics also leads 
to classical bubble production. In those works, large field excursions are generated by bubble collisions, which in turn result in the classical 
creation of new bubbles in minima lying below those of the parent bubbles. However, 
unlike the mechanism discussed in the present paper, the effect found in \cite{Easther:2009ft,Giblin:2010bd} operates only in models featuring a sequence of at least two false vacua arranged in a hierarchical structure.
Finally, let us also mention other studies that are somewhat similar in spirit, where phase transitions are triggered by various types of 
nucleation sites~\cite{Hosotani:1982ii,Preskill:1992ck},
such as neutron stars
\cite{Balkin:2021wea,Balkin:2021zfd}, black holes~\cite{Hiscock:1987hn,Gregory:2013hja,Mukaida:2017bgd,Shkerin:2021zbf}, domain walls \cite{Blasi:2022woz,Blasi:2023rqi,Agrawal:2023cgp}, cosmic strings \cite{Yajnik:1986tg,Blasi:2024mtc}, and monopoles
 \cite{Steinhardt:1981ec,Kumar:2010mv,Agrawal:2022hnf}.

The paper is structured as follows: in Section~\ref{sec:review} we briefly review dense boson stars, focusing on their collapse according to the self-similar solution found by \cite{Levkov:2016rkk}. In Section~\ref{sec:pt-from-bosenova}, we present the main result of this work and show for an explicit potential, how the boson star collapse can trigger the phase transition. In Section~\ref{sec:Pheno}, we comment on phenomenological applications of this mechanism. 
Finally in Section~\ref{sec:summary} we discuss our findings and provide with our concluding remarks. 
We also give a general discussion about the critical radius of a true vacuum bubble in Appendix~\ref{sec:Appendix A}.

\section{Boson stars: a review}
\label{sec:review}

Let us start by reviewing the basic properties of boson stars, which are chunks of Bose-Einstein condensate bound by 
gravity or attractive self-interactions.
Some of the discussions here follow those of
\cite{Guth:2014hsa,Levkov:2016rkk,Schiappacasse:2017ham}.

\subsection{Stable and unstable branches}

We consider a system at high occupancy, and describe it with a classical field theory of a real scalar coupled to gravity,
\bea
{\cal L}=\sqrt{-g}\l[ -\frac{1}{2} g^{\mu\nu} \partial_\mu \phi \, \partial_\nu\phi-V(\phi)\r],
\label{eq:Lag}
\eea
with a potential of the form,
\bea
\label{eq:gen-potential}
V(\phi)=\frac{1}{2}m^2 \phi^2 - m^2 f^2 \sum_{n=2}^{\infty} g_n\frac{(-1)^n}{(2n)!}\l(\frac{\phi}{f}\r)^{2n}.
\eea
Here $m$ is the mass, $g_n$ are dimensionless constants, and $f$ is a mass scale that serves as the decay constant in the case of axions.
In the nonrelativistic regime, it is convenient to rewrite $\phi$ as
\begin{equation}
\label{eq:subst-gpp}
    \phi (\bd{x},t) = \frac{f}{\sqrt{2 }}\left[ \psi(\bd{x},t)e^{-i m t} + \ro{c.c.} \right]\,,
\end{equation}
in terms of a complex field~$\psi$ whose time dependence is slow compared to the oscillation period $T=  2\pi/m$. 
Then the terms in the potential can be expressed as
\bea
\left( \frac{\phi}{f} \right)^{2 n}
=\frac{(2  n) !}{(n!)^2} \l(\frac{\psi \psi^*}{2}\r)^n+ (\hbox{oscillating terms}),
\eea
where the oscillating terms contain powers of 
$ e^{im t}$ and thus can be ignored if we average the 
field evolution over an oscillation period. 
We further take the metric as 
\begin{equation}
 ds^2 = -(1 + 2 \Phi) dt^2 + (1 - 2 \Phi) d\bd{x}^2,
\end{equation}
and use the weak gravitational field ($\abs{\Phi} \ll 1$)
and nonrelativistic ($\abs{\partial_\mu }\ll m$) approximations.
Then the Klein--Gordon equation,
$ \nabla_\mu \nabla^\mu \phi = d V / d \phi$,
and the time-time component of the Einstein's equation reduce to the Gross--Pitaevskii--Poisson (GPP) equations,
\begin{equation}
 \begin{split}
\label{eq:GPP}
    i\partial_{t}\psi = & -\frac{\partial_i \partial_i \psi}{2 m} + m\left(\Phi - \frac{g_{2}|\psi|^{2}}{8}\right)\psi \,,
\\
    \partial_i \partial_i \Phi = & \frac{4\pi \rho}{M_{p}^{2}}.
 \end{split}
\end{equation}
Here $M_p = G^{-1/2}$ is the Planck mass,
and $\rho = m^2 f^2 \abs{\psi}^2$ is the mass density.
Latin letters stand for spatial indices, and the sum over repeated spatial indices is implied irrespective of their positions.
Note that we have also made a small $\abs{\psi}$ assumption and dropped terms with higher powers of the field than those displayed.

Boson star solutions are obtained by numerically solving the GPP equations, but their basic properties can be understood by energy considerations using the Hamiltonian:\footnote{(\ref{eq:GPP}) arise as Hamilton's equations by using that the conjugate momentum of $\psi$ is 
$\pi = (i/2) m f^2 \psi^*$ (similarly for $\psi^*$ and $\pi^*$), 
and writing the Hamiltonian as 
$H = (H [\psi, \pi] + \ro{c.c.})/2$.}
\begin{equation}
 H = m^2 f^2 \int d^3 x 
\left( 
\frac{\partial_i \psi \partial_i \psi^*}{2 m^2}
- \frac{g_2}{16}\abs{\psi}^4
 \right)
- \frac{m^4 f^4}{2 M_p^2}
\int d^3 x \int d^3 x'
\frac{\abs{\psi(\bd{x})}^2 \abs{\psi(\bd{x}')}^2}{\abs{\bd{x} - \bd{x}'}}.
\label{eq:hamilt}
\end{equation}
The two terms in the parentheses correspond respectively to the 
gradient energy and the self-interaction.
The last term is the gravitational binding energy, for which 
we just showed the spatial variable of~$\psi$. 

Hereafter we suppose $g_2 > 0$, 
namely, an attractive quartic self-interaction
(which is the case for axion-like fields.)
Let us also consider a spherically symmetric ansatz with a characteristic radius~$R$, and an approximately constant field value inside this radius:
\begin{equation}
 \abs{\psi } \sim
 \begin{dcases}
  \abs{\psi}_0 \, \, (\ro{const.})
  & \mathrm{for}\, \, \,  
r \ll R, \\
  0 
  & \mathrm{for}\, \, \,  
r \gg R, \\
 \end{dcases}
\qquad
\abs{\partial_r \psi} \sim \frac{\abs{\psi}}{R},
\label{eq:star-ansatz}
\end{equation}
where $r = \abs{\bd{x}}$.
Then the integrals in (\ref{eq:hamilt}) can be performed to yield\footnote{For exact star solutions of the GPP equations, the terms in the Hamiltonian~(\ref{eq:star-H}) can be corrected by order-unity factors, whose values also depend on the detailed definitions of the star radius and mass~\cite{Ruffini:1969qy,Membrado:1989ke,Visinelli:2017ooc}. 
However the approximation~(\ref{eq:star-ansatz}) is good enough for order-of-magnitude estimates.}
\begin{equation}
 H \sim \frac{1}{2} \frac{M}{m^2 R^2}
- \frac{3 }{64 \pi } \frac{g_2 M^2}{m^2 f^2 R^3}
- \frac{3 }{5} \frac{M^2}{M_p^2 R},
\label{eq:star-H}
\end{equation}
where we have written the boson star mass as
$M = (4 \pi / 3) R^3 \rho_0$,
with the central density
$\rho_0 = m^2 f^2 \abs{\psi}_0^2$.
The Hamiltonian has two extrema with respect to the radius, 
a maximum at $R = R_-$ and a minimum at $R = R_+$ where
\begin{equation}
 R_{\pm}=
\frac{5}{6} \frac{M_p^2}{m^2 M}
\l(1\pm \sqrt{1 - \frac{27 }{80 \pi }\frac{g_2 m^2M^2}{f^2 M_p^2}} \r), 
\label{eq:Rpm}
\end{equation}
given that the star mass satisfies
\begin{equation}
M <  \sqrt{\frac{80 \pi }{27}}
\frac{f M_p}{\sqrt{g_2} m}.
\label{eq:max-mass}
\end{equation}
On the stable solution, $R = R_+$, the gravitational attraction balances the gradient pressure. On the other hand on $R= R_-$,
the gradient pressure balances the attractive self-interaction in an unstable equilibrium.
In the context of axion stars, these stable and unstable branches are also referred to as the dilute and dense branches, respectively.

\subsection{Boson star collapse}

We now focus on star configurations in the vicinity of the unstable branch, where gravity can be ignored. 
Then the GPP equations~(\ref{eq:GPP}) reduce to a 
nonlinear Schr{\"o}dinger-type equation, which has been well studied in condensed matter systems~\cite{Zakharov-Kuznetsov}, 
\bea
 i\partial_{t}\psi = -\frac{\partial_i \partial_i \psi}{2 m} - \frac{ m g_2|\psi|^{2}}{8}\psi. 
\label{sec:schrodinger}
\eea
As noted in \cite{Zakharov-Kuznetsov,Levkov:2016rkk}, this equation is invariant under the scaling,
\bea
\psi(\bd{x},t)\to \gamma \psi (\gamma \bd{x}, \gamma^2 t) e^{i \alpha}
\eea
with $\gamma, \alpha \in \mathbb{R}$. 
It thus has a scale-invariant solution of the form,
\bea
\label{eq:from-psi}
\psi = (-m t)^{-i w}\frac{\chi (y) }{\sqrt{g_2} m  r},~~y \equiv r \sqrt{\frac{m}{-t}},
\eea
where $w$ is a real constant, and we have set $t=0$ as the moment 
when the solution collapses to a singularity.
The function~$\chi(y)$ satisfies the equation
\bea
\label{eq:tkachev}
-2 w \chi +i y \chi'+\chi''+\frac{|\chi|^2 \chi}{4 y^2}=0,
\eea
where a prime denotes a $y$~derivative.

With the scale-invariant solution, the Hamiltonian~(\ref{eq:hamilt})
is written as (note we are ignoring gravity),
\bea
 H = \frac{2 \pi f^2 }{g_2 \sqrt{-m^3 t}}
\int_0^{\infty} 
\frac{d y }{y^2}
\left(
\abs{ \chi - y \chi' }^2
- \frac{\abs{\chi}^4}{8}
\right).
\eea
To make this integral converge, let us first require $\chi$ to vanish at the origin, $\chi(y \to 0) = 0$.
On the other hand at large~$y$, 
the general solution of (\ref{eq:tkachev}) asymptotes to the form (if $w \neq 0$),
\bea
\chi (y\to \infty ) = A y^{-2i w}+ B e^{-i y^2/2} y^{2 iw -1},
\eea
where $A$ and $B$ are complex constants.
The second term should vanish for the Hamiltonian to be finite. 
Numerically solving (\ref{eq:tkachev}) under these boundary conditions fix the parameters as $w\simeq 0.54$ and $\abs{A} \simeq2.85$~\cite{Levkov:2016rkk}.
Moreover, the absolute value of $\chi$ thus obtained is
well approximated by the following piece-wise function,
\begin{equation}
\label{eq:attractor-app}
|\chi |\simeq 
 \begin{dcases}
3.93 \,  y
  & \mathrm{for}\, \, \,  
y \ll 1, \\
2.85
  & \mathrm{for}\, \, \,  
y \gg 1, \\
 \end{dcases}
\end{equation}
or equivalently,
\begin{equation}
|\psi|\simeq 
 \begin{dcases}
\frac{3.93 }{\sqrt{-g_2 m t}}
  & \mathrm{for}\, \, \,  
m r \ll \sqrt{-m t}, \\
\frac{2.85}{\sqrt{g_2} m r}
  & \mathrm{for}\, \, \,  
m r \gg \sqrt{-m t}. \\
 \end{dcases}
\label{eq:axistar-collapse}
\end{equation}
If one extrapolates the two limiting expressions, they cross at $y \simeq 0.73$. 
Calling the region inside this $y$~value the core of the boson star, 
the core radius and the inner density are written as
\begin{equation}
 R \simeq 0.73 \sqrt{ \frac{-t}{m} },
\quad
 \rho_0 = m^2 f^2 \abs{\psi}_{0}^2
\simeq 15 \frac{m f^2}{- g_2 t}.
\label{eq:R-rho}
\end{equation}
Here we used the subscript~$0$ to denote quantities at $r \ll R$. 
Thus we see that the scale-invariant solution describes 
a star with a collapsing core, which has a uniform density growing with the collapse as
$\rho_0 \propto R^{-2}$.
In particular, the core mass and radius are related via
\begin{equation}
 M = \frac{4 \pi}{3} \rho_0 R^3 
\simeq 34 \frac{f^2}{g_2} R.
\label{eq:2.21}
\end{equation}
This relation matches with that of the 
unstable branch $R = R_-$ (cf. (\ref{eq:Rpm})) in the small~$M$ limit,
up to order-unity numerical factors. 

Two comments are in order here. 
First, it is the star core whose mass-radius relation takes a similar form as that of the unstable branch. 
For the whole star, since it acquires a non-uniform density during the collapse, the discussion below~(\ref{eq:star-ansatz}) no longer applies. 
Moreover, considering the scale-invariant profile~(\ref{eq:axistar-collapse}) to be cut off at the star radius~$R_{\rm{star}}$ ($\gg R$), the total mass of the star is much larger than the core mass:
$M_{\rm{star}} \simeq (3 R_{\rm{star}} / R) M \gg M$.
Second, it was demonstrated in \cite{Levkov:2016rkk} that the scale-invariant solution is an attractor. 
This indicates that a sufficiently dense boson star is dragged towards the scale-invariant core collapse.

Note that the above discussion holds in spacetime regions where the system is nonrelativistic, i.e. $-mt ,\, mr \gg 1$.
Moreover, the Schr{\"o}dinger equation (\ref{sec:schrodinger}) neglects
gravity, as well as sextic and higher self-interactions; 
the latter actually yields terms of 
$g_n \abs{\psi}^{2 n-2}$ inside the parentheses in the first line of 
the GPP equations~(\ref{eq:GPP}).
The Schr{\"o}dinger equation is thus valid for
$ g_2 \abs{\psi}^2 \gg \abs{\Phi}, \, \abs{g_n} \abs{\psi}^{2 n-2}$.
Imposing these conditions on the core and hence using 
$r \sim R$, $\abs{\psi} \sim \abs{\psi}_0$, and also estimating the gravitational potential as $\Phi \sim - G M / R$,
one finds that the scale-invariant solution is valid while the following conditions hold,
\begin{equation}
\frac{g_2 M_p^2}{f^2} \gg 
 -mt \gg
1, \, 
\left( \frac{\abs{g_n}}{g_2^{n-1}} \right)^{\frac{1}{n-2}},
\label{eq:break}
\end{equation}
for all $n \geq 3$. 
The first inequality represents the condition for negligible gravity; 
this combined with (\ref{eq:R-rho}) and (\ref{eq:2.21}) guarantees the star mass limit~(\ref{eq:max-mass}) to be satisfied.
One also sees from comparing the lower and upper limits on $-mt$
that the quartic coupling needs to satisfy 
\begin{equation}
 g_2 \gg \frac{f^2}{M_p^2},
\label{eq:g2cond}
\end{equation}
otherwise the scale-invariant solution would not exist at all.\footnote{(\ref{eq:g2cond}) roughly corresponds to requiring $R_- $ to be larger than the Schwarzschild radius, so that stars can reach the unstable branch before collapsing into black holes.}

The second inequality in (\ref{eq:break}) is for a nonrelativistic evolution, and negligible sextic and higher self-interactions. 
These eventually break down as the collapse proceeds ($ t \to 0^-$).
For the case of an axion-like potential, 
the scale-invariant collapse is followed by a ``bosenova'' which emits
relativistic axions and hence lowers the central density of the star~\cite{Levkov:2016rkk}.
The diluted star then collapses again along the unstable branch, and repeats the whole process. 

Lastly, we remark that we have used classical field theory to describe the star by assuming the system to be at high occupancy. 
To verify this assumption, let us evaluate the characteristic occupation number within the core as 
\begin{equation}
 \mathcal{N} \sim \frac{\rho_0}{m} \left(\frac{1 }{m v} \right)^3.
\end{equation}
Here $v$ is the typical velocity of the bosons, which we estimate from the gradient energy of the core (cf. first term in (\ref{eq:star-H})) via
$M/2 m^2 R^2 \sim M v^2 / 2$. This gives\footnote{On the stable branch, the expression here reduces to the virial velocity $v \sim \sqrt{G M / R_+}$ at small~$M$.}
\begin{equation}
 v \sim \frac{1}{m R}, 
\quad
 \mathcal{N} \sim \frac{M}{m}.
\label{eq:v}
\end{equation}
Hence during the core collapse, the occupation number decreases with the core mass. 
By extrapolating the scale-invariant solution~(\ref{eq:R-rho}) up until when the system becomes relativistic,
and using an asterisk to denote quantities at this time, 
i.e., $-m t_* = 1$,
one obtains $v_* \sim 1$ and $\mathcal{N}_* \sim f^2 / g_2 m^2$. 
The classical field description is valid until~$t_*$ if $\mathcal{N}_* \gg 1$, or equivalently,
\begin{equation}
 g_2 \ll \frac{f^2}{m^2}.
\label{eq:high-occu}
\end{equation}

\section{Phase transitions from star explosions}
\label{sec:pt-from-bosenova}

We explore the possibility that a collapsing boson star core triggers a cosmological phase transition. 
So far the collapse of boson stars has mainly been studied for trigonometric potentials. However, as was shown in the previous section, the essential ingredient for the collapse instability is a negative quartic coupling providing an attractive interaction. During
the  collapse  the energy density in the core of the boson star 
reaches very large values and the same is true for the classical 
field values.
Hence if the scalar field potential possesses non-degenerate vacua, the large field values within the star core can potentially
trigger a phase transition.
In the following we consider a classical ``jumping'' over the potential barrier induced by a star collapse, and discuss whether it creates a 
super-critical field configuration which subsequently starts to expand.

\subsection{A toy model}

We start with a toy potential:
\begin{equation}
\label{eq:potential}
    V(\phi) = \frac{m^{2}}{2!}\phi^{2} - \frac{\lambda}{4!}\phi^{4} + \frac{\epsilon}{6!}\phi^{6} \,,
\end{equation}
where $m$, $\lambda$, and $\epsilon$ are positive parameters with mass dimensions $1$, $0$, and $-2$, respectively.
(This corresponds to taking 
$g_2 = \lambda f^2 / m^2 $, $g_3 = \epsilon f^4 / m^2 $, and $g_4 = g_5 = \cdots = 0$
in the generic potential~(\ref{eq:gen-potential}).)
We impose the conditions~(\ref{eq:g2cond}) and (\ref{eq:high-occu})
for the existence of the unstable star branch, which in the current case translate into
\begin{equation}
\frac{m^2}{M_p^2} \ll \lambda \ll 1 .
\label{eq:lambda-cond}
\end{equation}

It is convenient to introduce dimensionless quantities,
\begin{equation}
\tilde{x}^\mu = m x^\mu,
\quad
\tilde{\phi} = \frac{\sqrt{\lambda}}{m} \phi, 
\quad
\delta = \frac{\epsilon \, m^2 }{\lambda^2}.
\label{eq:d-less}
\end{equation}
Then, ignoring gravity, the action is written as
\bea
S = \frac{1}{\lambda} \int d^{4}\tilde x \left( 
-\frac{1}{2} \eta^{\mu\nu} 
\frac{\partial \tilde{\phi}}{\partial \tilde{x}^{\mu}}
\frac{\partial \tilde{\phi}}{\partial \tilde{x}^{\nu}}
- \tilde{V} (\tilde{\phi}) \right),
\quad
\tilde{V} (\tilde{\phi}) = 
  \frac{\tilde \phi^{2}}{2} - \frac{\tilde \phi^{4}}{4!} 
 + \delta \frac{ \tilde \phi^{6}}{6!}
= \frac{\lambda }{m^4} V(\phi).
\eea
The classical evolution of the system thus depends only on $\delta$. 
Supposing $\delta < 5/8 = 0.625$, 
then the potential has a false vacuum at $\tilde{\phi}=0$ and true vacua at $\tilde{\phi}^2 = \tilde{\phi}_+^2$, separated by potential barriers 
peaked at $\tilde{\phi}^2 = \tilde{\phi}_-^2$, where
\begin{equation}
 \tilde{\phi}_{\pm}^2 = \frac{10}{\delta}
\left( 1 \pm \sqrt{1 - \frac{6 \delta }{5}} \right).
\end{equation}

We suppose that the system is initially trapped in the false vacuum, and that the probability for a quantum tunneling to the true vacuum is negligibly tiny. 
We further consider a boson star configuration around the false vacuum, that is sufficiently dense such that the quartic interaction balances the gradient pressure, but not dense enough for the sextic interaction to be relevant. 
Then as was shown in the previous section, 
we expect the star core to collapse 
along the usual unstable branch~(\ref{eq:axistar-collapse}),
up to the moments when the field value becomes large and the sextic interaction
can no longer be neglected, and/or the evolution becomes relativistic.

In order to test whether ``jumping'' over the barrier can happen, we numerically solved the classical equation of motion without gravity,
$ \partial_{\tilde{\mu}} \partial^{\tilde{\mu}} \tilde{\phi} = d \tilde{V} / d \tilde{\phi}$,
on a lattice.
We have implemented the simulation focusing only on spherically symmetric 
configurations, discretizing
space and time. The radial coordinate was discretized   with a  uniform grid with $\Delta \tilde{r} = 0.1$, and the time evolution was accounted for 
using the fourth order Runge-Kutta method with a fixed time step of $\Delta \tilde{t} = 10^{-4}$ (we have tested that the solution obtained with these discretizations is stable by comparing it with smaller values of $\Delta \tilde r,\, \Delta \tilde t$). At the boundary of the box the Neumann condition was imposed with $\d_{\tilde{r}} \tilde{\phi}=0$,\footnote{With this condition, outgoing waves can be reflected at the boundary, however we used a sufficiently large box such that the main core evolution concludes before the reflected waves reach the center.}
and likewise at the origin for regularity.
In order to reduce the time of the numerical integration we have used as initial conditions the attractor solution for collapsing stars in Eq.~\ref{eq:axistar-collapse}. 
More specifically, we first solved for $\chi$ as explained above
Eq.~\ref{eq:attractor-app}, then used the result to set the initial condition for $\phi$ for the subsequent Klein--Gordon calculation.
For these simulations we have tested field configurations with initial field values at the center of 
$\abs{\tilde{\phi}_{\ro{ini}}} \gtrsim 3 \times 10^{-2}$,
which collapse following the
attractor solution\footnote{On the unstable branch the field value within the star core fulfills 
$\abs{\tilde{\phi}} \gtrsim m / (\sqrt{\lambda} M_p)$. 
Hence here we are supposing that 
$m / (\sqrt{\lambda} M_p) \lesssim 10^{-2}$
(see also (\ref{eq:lambda-cond})).} within a time interval of $\Delta \tilde{t} \sim \abs{\tilde{\phi}_{\ro{ini}}}^{-2} \lesssim 10^3$. 
As expected the results were independent of where we set the initial condition along the unstable branch. 
We have also tested other spherically symmetric 
initial configurations, and checked that the system eventually relax to the attractor 
solution, although in this case the simulation lasted much longer since extra time was required for reaching the attractor.

\begin{figure}[t]
\begin{center}
\includegraphics[scale=1.45]{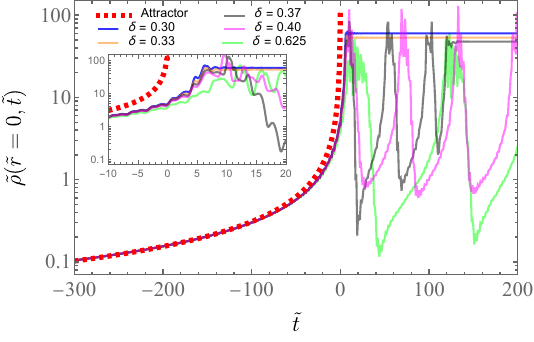}
\caption{$\tilde{\rho}\equiv \tilde{\phi}^2+\dot{\tilde{\phi}}^2 $
at the center of the star ($\tilde{r} = 0$), as a function of time. 
The numerical results are shown by the solid lines, whose colors denote different values of the sextic coupling~$\delta$.
The origin of time is chosen such that $\tilde{t} = 0$ corresponds to the moment when the self-similar attractor~(\ref{eq:axistar-collapse}),  shown as the red dashed line, becomes singular.
The sub-panel shows a zoom-in of the region around $\tilde{t}  = 0$.
For the green line ($\delta = 0.625$) the vacua are degenerate. 
For $\delta \lesssim 0.37$, the field becomes stabilized in the true vacuum and a phase transition is triggered. 
All quantities are in dimensionless units defined in~(\ref{eq:d-less}).
\label{fig:rho-g}
}
\end{center}
\end{figure}

\begin{figure}[t]
\centering
\subfigure[$\delta = 0.3$]{%
  \includegraphics[width=.85\linewidth]{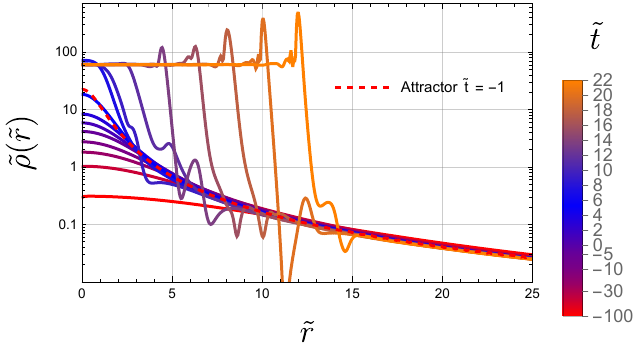}
  \label{fig:CB}}
\subfigure[$\delta = 0.5$]{%
  \includegraphics[width=.85\linewidth]{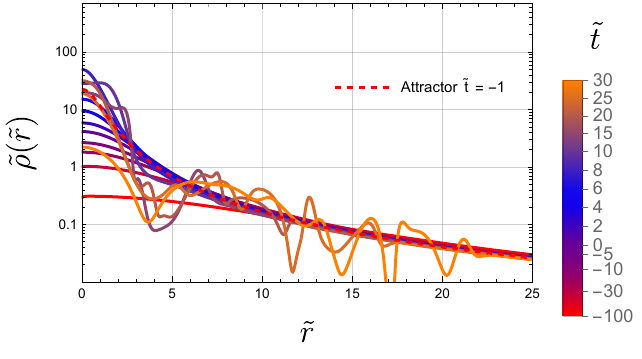}
  \label{fig:FB}}
 \caption{Snapshots of the radial profile of $\tilde{\rho}$. The line colors denote different times, as indicated by the tick values in the legends.
The red dashed line shows the self-similar profile~(\ref{eq:axistar-collapse}) at $\tilde{t} = -1$. 
($\tilde{t} = 0$ is when the self-similar profile develops a central singularity.)
The sextic coupling~$\delta$ is varied in the two panels.
In both panels the star core initially undergoes a self-similar collapse. 
In the upper panel a supercritical bubble forms in the core, and begins to expand.
In the lower panel a subcritical bubble forms, 
whose central field value turns to decrease as outgoing waves are emitted.
All quantities are in dimensionless units defined in~(\ref{eq:d-less}).}
\label{fig:ftvsr}
\vspace{\baselineskip}
\end{figure}

We report the numerical results
in Fig.~\ref{fig:rho-g} and Fig.~\ref{fig:ftvsr}, 
by plotting the quantity 
\begin{equation}
 \tilde{\rho} \equiv \tilde\phi^2+(\dot {\tilde\phi})^2,
\end{equation}
where the overdot denotes a $\tilde{t}$-derivative.
It is convenient to use $\tilde{\rho}$ as a proxy of the order parameter since, unlike the field~$\tilde\phi$, 
its oscillatory component with pre-factors $e^{\pm 2 i \tilde{t}}$ in the nonrelativistic regime is negligible (see Eq.~\ref{eq:subst-gpp}).
When this becomes as large as 
$\tilde{\rho} \gtrsim \tilde{V} (\tilde{\phi}_-) \sim 1$
(note that $\sqrt{6} < \tilde{\phi}_- < 2 \sqrt{2}$ for $0 < \delta < 5/8$), 
the field starts ``seeing'' the other side of the potential barrier.  

In Fig.~\ref{fig:rho-g} we show the time evolution of 
$\tilde{\rho}$ at the star center $\tilde{r} = 0$, where solid lines with different colors denote different values of~$\delta$. 
All the lines initially follow the self-similar collapsing solution~(\ref{eq:axistar-collapse}) which is represented by the red dashed line.
The origin of time is chosen such that $\tilde{t} = 0 $ corresponds to the moment when the self-similar collapse develops a central singularity.\footnote{We set the clock when the field evolution is on the self-similar attractor,
by defining $\tilde t=-300$ in terms of the central densities via
$\tilde\rho (-300)=\tilde \rho_{\rm att}(-300)$,
with the right-hand side obtained using~(\ref{eq:axistar-collapse}).}
The deviation from the self-similar collapse becomes substantial at  $\tilde{t} \sim -1$, and the subsequent core evolution depends on the value of~$\delta$. 
For $\delta$ not much smaller than~$0.625$, the vacua are nearly degenerate; for such cases one sees that after some oscillations around the true vacuum, the central~$\tilde{\rho}$ significantly drops, then turns to increase again. This recurrent behavior is similar to the bosenova phenomenon for axion-like potentials. 
On the other hand for smaller~$\delta$, the star collapse ceases with 
$\tilde{\rho}$ reaching a constant value of~$\tilde{\phi}_+^2$, 
which indicates that the field has settled down in the true vacuum. 
We find that such steady transitions occur
for the values of the coupling:
\bea 
\label{eq:gbound}
\delta \lesssim 0.37 .
\eea

In Fig.~\ref{fig:ftvsr} we have plotted the radial profile of $\tilde{\rho} $ at various snapshots of time, with $\delta$ varied in the two panels.
In both panels, the star follows the self-similar collapse~(\ref{eq:axistar-collapse}) at $\tilde{t} < -1$.
(The red dashed line depicts the self-similar profile at $\tilde{t} = -1$.)
In the upper panel, where $\delta $ is chosen to be below the threshold~(\ref{eq:gbound}), one can clearly see that the collapsing core turns to expand as a true vacuum bubble.
On the other hand in the lower panel where $\delta$ is above the threshold, the central~$\tilde{\rho}$ turns to decrease after reaching a maximum value; this decay of the high-density core is accompanied by emission of outgoing waves. 

A few comments are in order.
First, we have not included gravity in our simulations since we focused on stars initially on the unstable branch where gravitational effects can be safely ignored.
For generic star solutions, while the field is deep inside the potential well around the false vacuum, the contributions to the energy of the star core from gravity and quartic self-interaction are estimated in a similar fashion as, 
respectively, the third and second terms of the Hamiltonian~(\ref{eq:star-H}).
The ratio of the two contributions is thus of 
\begin{equation}
 \frac{E_{\ro{gr}}}{E_{\ro{int}}}
\sim \frac{m^4 }{\lambda M_p^2} R^2, 
\end{equation}
which shows that gravity becomes less and less relevant as the system size~$R$ shrinks during a collapse, independently of whether the system  follows the scale-invariant solution.
(Hence subcritical bubble cores are not expected to collapse into black holes.)
On the other hand, after the field makes the transition to the true vacuum, gravity can become relevant as the vacuum bubble expands; however we expect our analysis to remain valid while
the bubble size is much smaller than the Hubble radius.

Second, we have not studied in depth the basin of attraction for the collapsing solution, and how spherically non-symmetric configurations evolve. 
However since the collapse is initially governed by the quartic interaction, 
the results in Ref.~\cite{Levkov:2016rkk} which partially demonstrated that the spherically symmetric collapsing solution is an attractor for axion stars, also apply to our case.
If the star becomes sufficiently spherically symmetric during the scale-invariant collapse, then it is reasonable to expect that the system remains nearly spherically symmetric also during the subsequent evolution.

Finally, we have relied on classical field analyses by considering the system to always be in the high occupancy regime. This is guaranteed by 
the condition, $\lambda \ll 1$ in (\ref{eq:lambda-cond}), up until the time when the core becomes relativistic. 
Even beyond this time, however, quantum corrections should be small for the following reasons:
In cases where a supercritical bubble is formed, the field value inside the bubble is at least of $m / \sqrt{\lambda}$; this is larger than quantum fluctuations which should be comparable to the 
inverse size of the initial bubble,~$m$. 
In cases with a subcritical bubble, the emission of outgoing waves bring the system back into the nonrelativistic regime, where the occupation number is larger. 
Hence for both cases, we expect the classical field description to be valid as long as $\lambda$ is sufficiently smaller than unity.

\subsection{Comparison with bounce solutions}

The existence of the upper bound 
on $\delta$ for triggering a transition can be understood 
from the following qualitative 
considerations. 

The attractor solution (\ref{eq:axistar-collapse}) breaks down at $t \sim t_* = -1/m$, when the core evolution becomes relativistic. 
The sextic interaction becomes nonnegligible at around the same time or later since $ g_3 / g_2^{2} = \delta < 5/8$, cf.~(\ref{eq:break}). 
The core radius at~$t_*$ is of
\begin{equation}
 R_* \sim \frac{1}{m},
\end{equation}
inside which the field's oscillation amplitude reaches the top of the potential barrier, i.e., 
$ f \abs{\psi_*}_0 \sim  \phi_-$.
We can hence crudely picture the core at this time as a bubble of true vacuum. 

A true vacuum bubble within the false vacuum 
expands only if its radius is 
large enough so that the driving force 
from the potential difference wins over the surface tension. 
If the two vacua are nearly degenerate such that
$V(0) - V(\phi_+) \ll V(\phi_-) - V(\phi_+)$
(which is the case for $\delta \simeq 5/8$),
then the wall of the bubble is thin. The 
critical radius~$R_B$ of such thin-wall bubbles can be found by extremizing the bubble energy,
\bea
E(R)=4\pi R^2 \sigma  -\frac{4\pi}{3} R^3 
\left\{ V (0) - V (\phi_+) \right\}
\eea
where $\sigma$ is the surface tension, as
\bea
R_B=\frac{2 \sigma }{ V (0) - V (\phi_+) }.
\label{eq:thinwall}
\eea
We thus expect that the true vacuum core expands and triggers a cosmological phase transition if its initial radius is larger than the critical radius,\footnote{Alternatively, one can set the condition to expand as the field value at the critical radius being larger than that of the bounce solution, 
$f \abs{\psi (R_B, t_*) } > \abs{\phi_B (R_B)}$. 
This also gives a similar bound on~$\delta$.}
\bea
\label{eq:cond-gener}
R_* > R_B.
\eea

\begin{figure}
\centering
\includegraphics[scale=1.25]{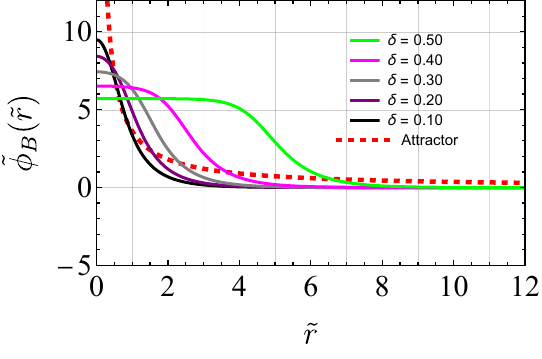}
\caption{Profiles of bounce solutions for various values of the coupling~$\delta$. 
The red dashed line shows the self-similar profile~(\ref{eq:axistar-collapse}) in the limit~$\tilde{t} \to 0^-$. 
The rough condition~(\ref{eq:cond-gener}) for supercriticality is satisfied for $\delta \lesssim 0.2$. 
All quantities are in dimensionless units defined in~(\ref{eq:d-less}).
\label{fig:bounces}}
\end{figure}

For generic bubbles which do not necessarily have thin walls, the critical field configuration 
of a spherically symmetric bubble, which we refer to as 
the bounce solution~$\phi_B (r)$,
can be found by solving the equation (see Appendix~\ref{sec:Appendix A} for more details),
\bea
\frac{d^2 \phi_B}{d r^2}+\frac{2}{r}\frac{d\phi_B}{dr}=\frac{d V (\phi_B)}{d\phi_B},
\eea
with boundary conditions,
\begin{equation}
\phi_B (0) \neq 
 \phi_B(r \rightarrow \infty ) =0\,, \quad 
\left. \frac{d \phi _B}{dr} \right|_{r = 0} =0\,.
\end{equation}
We have numerically obtained the bounce profiles, which we plot in Fig.~\ref{fig:bounces} for various values of~$\delta$. 
The exact radius of these solutions are ambiguous owing to their thick walls, hence we define the critical radius via
\bea
\label{eq:RBradius}
\phi_B(R_B)\equiv \frac{\phi_B(0)}{2},
\eea
as the radius where the field value drops by a factor of two compared to the origin.
In the plot, we also show by the red dashed curve
the dimensionless field's oscillation amplitude 
$\sqrt{2 \lambda} f \abs{\psi} / m$
for the self-similar expression~(\ref{eq:axistar-collapse}) extrapolated to  $\tilde{t} \to 0^-$.

One can read off from the plot that the condition~(\ref{eq:cond-gener}), i.e. $ \tilde{R}_B = m R_B \lesssim 1$, 
reduces to an upper bound on the coupling,
\bea
\label{eq:cond-naive}
\delta \lesssim 0.2.
\eea
For much larger values of~$\delta$, the critical radius $R_B$ becomes too large such that the true vacuum cores cannot expand. 
We should remark that this bound should only be taken as an indicative estimate since 
the definition of $R_B$ is somewhat arbitrary, 
and the discussion is based on 
a comparison between a static critical field configuration, and a time-dependent collapsing star solution which first needs to come to a halt before turning to expand.
However it qualitatively explains the existence of the upper bound on $\delta$ and 
is of the same order as the exact one~(\ref{eq:gbound}).\footnote{The threshold coupling $\delta=0.37$ in (\ref{eq:gbound}) gives 
$\tilde{R}_B \simeq 2.3$. On the other hand, the minimal critical radius is obtained in the limit $\delta \to 0$ as $\tilde{R}_B \simeq 0.56$.}

\subsection{Other potentials}

So far, our discussion was focused on the potential with a negative quartic and a positive sextic coupling, as given in Eq.~\ref{eq:potential}. One may naturally ask whether similar phenomena can occur for other types of potentials. Intuitively, this seems plausible, since the collapse is primarily driven by attractive self-interactions in the infrared (IR). 
Here we briefly comment on other potentials that might exhibit similar IR dynamics.

\subsubsection{Logarithmic potential}

Let us start by  considering  the potential:
\bea
\label{eq:log-pot}
V=\frac{m^2 \phi^2}{2}+\frac{\lambda\phi^4}{4} \log\frac{\phi^2}{\Lambda^2},
\eea
with $\lambda > 0$. 
We would like to emphasize that for small field values such that
$\phi^2 < \Lambda^2$, the product~$\lambda \log (\phi^2 / \Lambda^2) $ becomes negative. By considering this as an effective negative quartic coupling, one expects that the log potential induces a star collapse. 
However, unlike for genuine negative quartic potentials, we are not aware if an attractor solution exists. Hence to explore the parameter space we have to rely fully on numerical simulations, which are also
limited by the grid size and total time of the simulation. 

For numerical simulations it is convenient to  perform the field redefinition similarly to the previous section to obtain a dimensionless parameterization:
\begin{equation}
\begin{split}
& \tilde{x}^\mu = m x^\mu,
\quad
\tilde{\phi} = \frac{\sqrt{\lambda}}{m} \phi, 
\quad
\zeta = \frac{m^2}{\lambda \Lambda^2},
\\
& S = \frac{1}{\lambda} \int d^{4}\tilde x \left( 
-\frac{1}{2} \eta^{\mu\nu} 
\frac{\partial \tilde{\phi}}{\partial \tilde{x}^{\mu}}
\frac{\partial \tilde{\phi}}{\partial \tilde{x}^{\nu}}
- \frac{\tilde \phi^{2}}{2} - \frac{\tilde \phi^{4}}{4} 
\log (\zeta \tilde{\phi}^2)
\right).
\end{split}
\end{equation}
For $\zeta \lesssim 0.18$, the potential possesses true vacua at $\tilde{\phi} \neq 0$.
The bounce action, the bounce solution, and the bounce radius can be written in terms of dimensionless functions as follows:
\bea
S_B=\frac{1}{\lambda}\tilde S_B\l( \zeta \r),~~~
   \phi_B(r)= \frac{m}{\sqrt{\lambda}}\tilde \phi_B\l(\zeta, \tilde{r} \r),~~R_B= \frac{1}{m} \tilde R_B\l( \zeta \r).
\eea

In order to identify the parameter space where the
phase transition  can be triggered, 
we have performed numerical simulations imposing spherical symmetry, with initial configurations of the form
$\tilde \phi =\tilde \phi_c/(1+\tilde r^2/\tilde R_c^2)$ 
which is regular at the origin and yields a finite total energy.
Due to the time constraints on the simulations we have kept $\tilde \phi_c\gtrsim 0.03$. $\tilde R_c$ is the initial size of the core, which needs to be 
sufficiently large for the core to collapse. For example for $\tilde \phi_c=0.03$, one needs $\tilde R_c \gtrsim 60 $. 
With these simulations, we found that an expanding true vacuum bubble is formed from a core collapse for 
\bea
\zeta \lesssim 0.1,
\eea
with this result being independent of the initial field parameters.
We also tested initial conditions given by the attractor solution for the quartic potential in Eq.~\ref{eq:axistar-collapse}
(even though it is not a solution for the log potential), and observed phase transitions happening.
The existence of an upper limit on~$\zeta$ for a collapsing core to turn to expand can be understood by noting that, similar to the sextic potential, as $\zeta $ increases the false and true vacua become degenerate, and hence the critical radius becomes larger. 
However let us stress again that it is not known if an attractor solution exists for the logarithmic potential. And since we have only managed to test a small set of initial configurations, the results here 
can be considered only as a hint of the star-induced bubble nucleation, until more rigorous study is done.

\subsubsection{Cubic potential}
One may wonder whether similar effects can happen for a cubic potential, since this is the simplest renormalizable potential 
that can have a false vacuum:
\bea
V=\frac{m^2 \phi^2}{2}-k \frac{\phi^3}{3}+\lambda \frac{\phi^4}{4},
\eea
where $\lambda > 0$. 
In this case the cubic interaction makes the potential flatter than a quadratic, and hence may induce an attractive force.
(However it should also be noted that 
if one makes the substitution as in Eq.~\ref{eq:subst-gpp}, odd powers of $\phi$ only give rise to oscillatory terms.)
Since the only interaction, if any, that can trigger a star collapse
is cubic, the resulting behavior is expected to be quite different from the potentials studied in the previous sections.

Making the field redefinitions:
\begin{equation}
\begin{split}
& \tilde{x}^\mu = m x^\mu,
\quad
\tilde{\phi} = \frac{k}{m^2} \phi, 
\quad
\xi = \frac{\lambda m^2}{k^2},
\\
& S = \frac{m^2}{k^2} \int d^{4}\tilde x \left( 
-\frac{1}{2} \eta^{\mu\nu} 
\frac{\partial \tilde{\phi}}{\partial \tilde{x}^{\mu}}
\frac{\partial \tilde{\phi}}{\partial \tilde{x}^{\nu}}
-\frac{\tilde \phi^2}{2}+\frac{\tilde \phi^3}{3} - \xi \frac{\tilde \phi^4}{4}
\right),
\end{split}
\end{equation}
one sees that the dynamics of the system is controlled by $\xi$. 
The potential has a true vacuum at $\tilde{\phi} \neq 0$ for $\xi < 2/9 \simeq 0.22$.
We numerically computed spherically symmetric configurations, with initial profiles that decay at large radii as $\tilde{\phi} \propto \tilde{r}^{-2}$, with central field values 
$\tilde{ \phi}_{c}\gtrsim 0.05$ due to time limitations.
We observed star configurations collapsing and then turning to expand as true vacuum bubbles. 
Within the cases we tested, a critical bubble creation took place independently of the initial conditions under the condition,
\bea
\label{eq:cubic-bound}
 \xi \lesssim 0.1.
\eea
We emphasize that we only studied a small set of initial configurations, and moreover it is still unclear if a cubic interaction should give rise to an attractive force. Hence at this point the collapse and the subsequent expansion of stars for cubic potentials should only be considered as a conjecture.

\section{ Phenomenological  implications}
\label{sec:Pheno}

In the previous section, we introduced a mechanism by which the collapse of a boson star can trigger a phase transition. Until now, we have remained agnostic regarding the cosmological conditions required for this process. In what follows, we outline the basic criteria the system must fulfill for such a transition to occur,
by focusing on the sextic model~(\ref{eq:potential}).

\subsection{Transition to our universe}

First and foremost, the universe must initially reside in the false vacuum, which must be stable against quantum tunneling.
The tunneling rate, $\Gamma \propto \exp [-S_4]$, 
is controlled by the Euclidean action evaluated using an O(4)-invariant bounce solution (which is obtained in a similar fashion as the O(3)-invariant solution described in 
Appendix~\ref{sec:Appendix A}). 
The bounce action depends on the couplings as follows,
\bea
S_4=\frac{1}{\lambda}\tilde S_4 \l(\delta\r) > 0.
\eea
At the same time the production of the supercritical bubbles from the star collapse is controlled only by the $\delta$ parameter. So if we keep $\delta$ fixed and choose $\lambda$ small enough we can make the metastable state arbitrarily stable against tunneling.

Let us now assume that our current universe 
exists in the true vacuum of the system.\footnote{The sextic potential possesses two degenerate true vacua, and a collapsed core settles down in one of them. With multiple stars, half of them collapse into one vacuum, and the rest into the other vacuum, thus giving rise to domain walls (as is generally the case for phase transitions to discrete vacua.) Hence an energy bias between the vacua needs to be added to remove the walls. However here we show that transition-inducing stars do not form in the first place with the sextic potential.}
In this case, it follows from the observed value of the cosmological constant which is vanishingly small that the false vacuum must possess a constant contribution to the energy density, given by
\bea
\label{eq:vac-energy}
\begin{split}
V(0) - V(\phi_+)
&=\frac{m^4}{\lambda}\l[ \frac{(5+\sqrt{25-30 \delta})(5+\sqrt{25-30 \delta}-12 \delta)}{18 \delta^2}\r] \\
&= \frac{m^4}{\lambda}\l[ \frac{50}{9 \delta^2}-\frac{10}{\delta} +O(1) \r]
\end{split}
\eea
In the second line, we have expanded in the small $\delta$ limit ($\delta\ll 1$), and we remind the reader that $(0,\phi_-,\phi_+)$ are the field values in the false vacuum, local maximum of the potential and true minimum respectively. 
The presence of this vacuum energy imposes a stringent constraint on the star formation, as we describe below.

Boson stars can form in virialized halos through a
condensation~\cite{Kolb:1993zz,Levkov:2018kau,Eggemeier:2019jsu,Chen:2021oot} or by a gravothermal collapse~\cite{Ralegankar:2024zjd};
they can also form from the collapse of large-amplitude density fluctuations~\cite{Gorghetto:2024vnp}. In any cases, the formation of boson stars (or at least their host halos) typically takes place in a universe dominated by nonrelativistic matter. 
Hence one may imagine a scenario that leads to the star-induced phase transition as follows:
The $\phi$~particles are initially excited around the false vacuum, which eventually become nonrelativistic and dominate the universe. 
During this early matter-dominated epoch, the $\phi$~particles form boson stars through one of the above mechanisms. 
The stars then grow by accreting the surrounding particles~\cite{Chan:2022bkz,Dmitriev:2023ipv}, some of which reach the unstable branch, collapse, and explode as bubbles of true vacuum.

The above scenario requires an epoch dominated by the $\phi$~particles in the false vacuum. 
However their energy density, that is,
\bea
\rho_{\phi}\sim {\phi}^2 m^2,
\eea
cannot be larger than the height of the potential barrier separating the false and true vacua, otherwise the field does not remain trapped in the false vacuum. Hence,
\bea
\label{eq:vac-max}
\begin{split}
\rho_{\phi}< 
V(\phi_-) - V(0) &= \frac{m^4}{\lambda} \l[\frac{(\sqrt{25-30 \delta}-5)(5-12 \delta-\sqrt{25-30 \delta})}{18 \delta^2}\r] \\
&= \frac{m^4}{\lambda} \l[\frac{3}{2}+\frac{3 \delta}{10} + O(\delta^2)\r],
\end{split}
\eea
for $\delta\ll 1$.
Then using Eqs.~\ref{eq:vac-energy} and \ref{eq:vac-max}, as well as
the upper limit on~$\delta$ in Eq.~\ref{eq:gbound} for the true vacuum core to expand, 
we can conclude immediately that 
\bea
\rho_{\phi} < V(\phi_-) - V(0) < V(0) - V(\phi_+) .
\eea
Thus there cannot be a period of matter domination by the $\phi$~particles in the false vacuum, 
disallowing star formation in the first place. 
In other words, boson stars could form 
if the $\delta$-limit of Eq.~\ref{eq:gbound} is violated, 
but then their collapse would not give rise to expanding true vacuum bubbles.

The formulas above are derived assuming the sextic potential in Eq.~\ref{eq:potential}. 
However even for generic potentials, 
the requirement that the height of the barrier be larger than the energy difference between the vacua restricts the bounce solution to have a thin wall, 
and an accordingly large critical radius. 
Unless an even larger true vacuum bubble is formed from the star collapse, the main conclusion remains to hold.

We have also assumed a matter-dominated epoch driven by the same field 
producing the boson stars.  
If the stars or at least their host halos can form during a different epoch,
the previous discussion in principle does not apply.
For instance, one can imagine another matter component that is not bounded by $\phi$'s potential barrier to dominate the universe and induce structure formation. Alternatively, halos might form in a non-matter-dominated universe with the aid of additional long-range forces~\cite{Flores:2020drq}.

\subsection{Transition from our universe}

\begin{figure}
    \centering
    \includegraphics[width=0.5\linewidth]{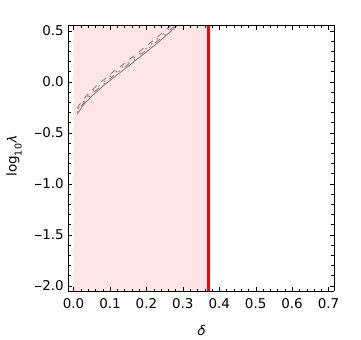}
    \caption{Parameter space where phase transitions can happen, in terms of the couplings $\delta$ and $\lambda$ for the sextic model. Below black lines the universe is stable against quantum tunneling up to today,
with the mass of $\phi$ varied as $m = 1\, \ro{GeV}$ (dashed), $10^2\, \ro{GeV}$ (dotted), $10^4\, \ro{GeV}$ (dot-dashed), and 
$10^6\, \ro{GeV}$ (solid).
In the red shaded area phase transitions can happen due to the boson star collapse.}
    \label{fig:lambda-g}
\end{figure}

Another possibility is that the observed universe lives in the false vacuum, i.e., it is metastable.  
If the energy of the true vacuum is negative, then the potential barrier height can be 
smaller than the energy difference between the vacua, and also be 
much larger than the absolute value of the false vacuum energy.
The $\phi$~particles then could have once dominated the universe (perhaps as dark matter) and formed dense boson stars.
In this case a phase transition leads to the destruction of our universe,
so we can constrain theories by requiring this not to happen, neither by collapsing stars nor by quantum tunneling.

The tunneling rate can be estimated as
\bea
\Gamma\sim \frac{1}{R_4^4}\l(\frac{S_{4}}{2\pi }\r)^2\exp\l[- S_{4}\r] = \frac{m^4}{\tilde R_4(\delta)^4}\l(\frac{\tilde S_{4}(\delta)}{2\pi \lambda}\r)^2\exp\l[-\frac{1}{\lambda}\tilde S_{4}(\delta)\r].
\eea
Here, $R_4 = \tilde{R}_4(\delta) / m$ is the radius of the O(4)-invariant bounce solution, defined in a similar way as in Eq.~\ref{eq:RBradius}. 

In Fig.~\ref{fig:lambda-g} we show the parameter space for metastability 
in the $\delta$-$\lambda$ plane. 
A quantum tunneling takes place by today, 
i.e., $\Gamma > \mathcal{H}^4$ where 
$\mathcal{H} \sim 10^{-33}\, \ro{eV}$ is the present-day Hubble rate,
on the left side of the black lines (whose exact position also depends on $\phi$'s mass).
On the other hand, a collapsing boson star triggers a phase transition on the left of the red line, cf. Eq.~\ref{eq:gbound}.
We can see that 
there is a large parameter space at $\lambda \ll 1$ where the otherwise perfectly fine metastable 
universe can be destroyed by 
a boson star collapse.
On the other hand at $\lambda \gtrsim 1$, not only that the quantum tunneling becomes important, but the classical field description can break down at the final stage of the star collapse, cf.~(\ref{eq:lambda-cond}).

\section{Summary \& Conclusion}
\label{sec:summary}

We conclude by summarizing the main results of our study. We investigated systems in which a scalar field is trapped in a false vacuum, with particular focus on potentials characterized by negative quartic interactions. 
It is well established that such a potential can lead to the formation of boson stars, which subsequently collapse if 
their densities exceed a critical value.
The collapse of the boson star follows an attractor solution, resulting in very large field values at the core of the star. 
We have demonstrated that configurations with sufficiently large field values can classically trigger a phase transition by creating regions of true vacuum at the core of the boson star. If these regions exceed the size of a critical bubble, they begin to expand, thereby initiating a first-order phase transition.

We have analyzed star collapse in detail for a model with $\phi^4$ and $\phi^6$ interactions, and identified the parameter window where 
a phase transition can be triggered through first-principle numerical 
simulations of spherically symmetric field configurations. 
We found phase transitions are induced for 
potential parameters that admit bounce solutions with thick walls, i.e., 
bubbles with relatively small radii that are comparable to the scale of the underlying microphysics.
This behavior is expected, as in such cases the critical bubble radius can be smaller than the core size of the collapsing star.

Interestingly, our numerical 
results reveal qualitatively similar behavior for other types of potentials—
specifically, logarithmic and cubic ones —that may also exhibit attractive interactions at infrared 
field values. However, unlike the case with a negative quartic coupling, no attractor solution is 
known for these potentials. 
To what extent the boson star collapse and subsequent phase transition in these scenarios depend on the initial conditions remains to be understood.

We have briefly discussed the phenomenological implications. 
In particular, we demonstrated that a false vacuum that is metastable against quantum tunneling and hence may seem 
to have a lifetime exceeding the age of the universe,
can actually be destabilized by a boson star collapse.
At this point there are  two distinct scenarios: (i)~our world lives in the true vacuum, or (ii)~our world lives in the false vacuum. Regarding the case~(i),
for the simplest toy scenario
we showed that boson stars formed in a matter-dominated universe in the false vacuum can only collapse into subcritical bubbles, and hence do not trigger a phase transition. 
It would be interesting to see if a similar situation also arises for other boson potentials and star formation scenarios.
For~(ii), the false vacuum energy is normalized to the observed value of the cosmological constant and is very small. 
In this case, boson stars can form and trigger a destruction of the otherwise metastable universe. 
The requirement that such dangerous stars do not form can be used to 
constrain cosmological models with multiple vacua, 
such as those with
multi-axion couplings or gravitational corrections generating superposed cosine potentials~\cite{Graham:2015cka,Choi:2015fiu,Kaplan:2015fuy,Higaki:2016yqk,Kobayashi:2018nzh,DiLuzio:2021gos}.

We also comment on gravitational waves emitted from the collision of bubbles (see for a recent review \cite{Athron:2023xlk}). The frequency is set by the initial distance between the bubbles upon nucleation, which is roughly the Hubble radius in conventional first-order phase transition scenarios. 
However in our mechanism, the bubble separation should be much smaller since it corresponds to the interstellar distance.
Hence for phase transitions happening at the same energy scale, 
we expect star-induced bubbles to produce gravitational waves with much higher frequencies than in conventional scenarios.
On top of this, there may be additional gravitational wave emission from mergers of boson stars~\cite{Croon:2018ftb,Chung-Jukko:2024hod},
and from fluctuations of the expanding bubble wall~\cite{Blum:2024hcs}.

Finally, we remark that although we have focused on classical transitions induced by collapsing stars, the large core density may also enhance the quantum tunneling rate. 
Developing a quantitative understanding of this effect is both interesting and worthwhile, and we leave this to future work.

\acknowledgments

We thank Motoo Suzuki for initial collaboration.
We are also thankful to 
Francescopaolo Lopez, Pranjal Ralegankar, Chen Sun, and Michael Zantedeschi
for helpful discussions. 
This work was supported in part by the European Union - NextGenerationEU through the PRIN Project ``Charting unexplored avenues in Dark Matter'' (20224JR28W), and INFN initiatives TAsP and APINE.
T.K. also acknowledges support from JSPS KAKENHI (JP22K03595).

\appendix
\section{O(3)-invariant bounce solution}
\label{sec:Appendix A}

In this appendix we discuss the critical radius of a true vacuum bubble in terms of an O(3)-invariant bounce solution. 
These  static field configurations correspond to 
a saddle point of the  
energy functional, and under 
small perturbations can either collapse or expand.
Our discussion is purely classical, however 
the computation is essentially the same as for 
critical bubble formation at finite temperatures due to thermal jumps~\cite{Linde:1980tt}.

We consider a real scalar field with a Lagrangian of the form~(\ref{eq:Lag}) in flat space, with a potential~$V(\phi)$ possessing two non-degenerate minima.
We set the field value at the false vacuum to $\phi = 0$, without loss of generality.
We restrict ourselves to static and spherically symmetric bubble solutions, i.e.,
$\phi = \phi_B (r)$ with $r = \abs{\bd{x}}$.
Then the field's equation of motion takes the form,
\begin{equation}
 \frac{d^2 \phi_B}{d r^2}+ \frac{2}{r} \frac{d \phi_B }{dr} = \frac{dV(\phi_B)}{d \phi_B}.
\label{eq:O3eom}
\end{equation}
In order to model a bubble of true vacuum within the false vacuum, we also impose the following boundary conditions,
\begin{equation}
\phi_B (0) \neq 
 \phi_B(r \rightarrow \infty ) =0\,, \quad 
\left. \frac{d \phi _B}{dr} \right|_{r = 0} =0\,,
\label{eq:boundary}
\end{equation}
where the second condition is for regularity at the origin. 

Interpreting the radial variable~$r$ as time, then the equation~(\ref{eq:O3eom}) describes the motion of a particle moving along an inverted potential~$-V$, being also subject to a frictional force of $(2/r) (d \phi / dr)$. 
The boundary conditions further impose that the particle is initially ($r = 0$) at rest, then subsequently starts to roll and reaches~$\phi = 0$ in the asymptotic future ($r \to \infty$). Here, since $\phi = 0$ is a local maximum of the inverted potential, to reach it the particle needs to be initially located in the potential well around the true vacuum. 
Moreover, due to the presence of the frictional force, the initial position needs to fulfill 
$V(\phi_B (0)) < V(0)$. 
We refer to $\phi_B (r)$ as the ``bounce'' solution. 
This can be numerically obtained for generic potentials using a shooting method, or by using existing codes such as FindBounce~\cite{Guada:2020xnz} and CosmoTransitions~\cite{Wainwright:2011kj}.

The  bounce solution is unstable under deformations. For example we can look at the following one-parameter field configurations
$ \phi_\kappa (r) = \phi_B ( \kappa r )$ with $\kappa > 0$
(hence $\phi_{\kappa = 1} = \phi_B$)~\cite{Coleman:1985rnk}.
Noting that the bounce solution extremizes the Hamiltonian for~$\phi$,
\begin{equation}
 H[\phi] = 4 \pi \int^{\infty}_0 dr \, r^2
\left[
\frac{1}{2} \left( \frac{d \phi }{dr} \right)^2 + V
\right],
\end{equation}
and thus $( d H[\phi_\kappa] / d \kappa )_{\kappa = 1} = 0$, 
one can check the instability:
\begin{equation}
 \left(
 \frac{d^2 H[\phi_\kappa ]}{d \kappa^2}
\right)_{\kappa = 1}
= - 4 \pi 
\int^{\infty}_0 dr \, r^2
\left( \frac{d \phi_B }{dr} \right)^2 
< 0.
\end{equation}
This shows that a bubble larger (smaller) than the bounce solution expands (shrinks). 

In cases where the true and false vacua are almost degenerate, the bounce solution exhibits a quick transition between the two vacua at a sufficiently large~$r$ such that the friction is negligible during the transition. This describes a bubble with a ``thin wall,'' whose radius can be computed as in (\ref{eq:thinwall}), with the surface tension $\sigma$ 
being a functional of $V$. The radius of a thin-wall bounce solutions gives the critical radius, beyond which bubbles begin to expand. 
However in general cases, bubbles can have thick walls, then whether they expand or shrink also depends on the shape of the field profile.
In the main text, we introduced a generalized critical radius as (\ref{eq:RBradius}), as a rough guide for determining the fate of bubbles. 

\bibliographystyle{JHEP}
\bibliography{biblio}

@article{Levkov:2016rkk,
    author = "Levkov, D. G. and Panin, A. G. and Tkachev, I. I.",
    title = "{Relativistic axions from collapsing Bose stars}",
    eprint = "1609.03611",
    archivePrefix = "arXiv",
    primaryClass = "astro-ph.CO",
    reportNumber = "INR-TH-2016-034",
    doi = "10.1103/PhysRevLett.118.011301",
    journal = "Phys. Rev. Lett.",
    volume = "118",
    number = "1",
    pages = "011301",
    year = "2017"
}

@article{Ruffini:1969qy,
    author = "Ruffini, Remo and Bonazzola, Silvano",
    title = "{Systems of selfgravitating particles in general relativity and the concept of an equation of state}",
    doi = "10.1103/PhysRev.187.1767",
    journal = "Phys. Rev.",
    volume = "187",
    pages = "1767--1783",
    year = "1969"
}

@article{Balkin:2021zfd,
    author = "Balkin, Reuven and Serra, Javi and Springmann, Konstantin and Stelzl, Stefan and Weiler, Andreas",
    title = "{Density induced vacuum instability}",
    eprint = "2105.13354",
    archivePrefix = "arXiv",
    primaryClass = "hep-ph",
    reportNumber = "TUM-HEP-1308/20",
    doi = "10.21468/SciPostPhys.14.4.071",
    journal = "SciPost Phys.",
    volume = "14",
    number = "4",
    pages = "071",
    year = "2023"
}

@article{Chavanis:2025qcg,
    author = "Chavanis, Pierre-Henri",
    title = "{A review of basic results on the Bose{\textendash}Einstein condensate dark matter model}",
    doi = "10.3389/fspas.2025.1538434",
    journal = "Front. Astron. Space Sci.",
    volume = "12",
    pages = "1538434",
    year = "2025"
}

@article{Chavanis:2017loo,
    author = "Chavanis, Pierre-Henri",
    title = "{Phase transitions between dilute and dense axion stars}",
    eprint = "1710.06268",
    archivePrefix = "arXiv",
    primaryClass = "gr-qc",
    doi = "10.1103/PhysRevD.98.023009",
    journal = "Phys. Rev. D",
    volume = "98",
    number = "2",
    pages = "023009",
    year = "2018"
}

@article{Balkin:2021wea,
    author = "Balkin, Reuven and Serra, Javi and Springmann, Konstantin and Stelzl, Stefan and Weiler, Andreas",
    title = "{Runaway relaxion from finite density}",
    eprint = "2106.11320",
    archivePrefix = "arXiv",
    primaryClass = "hep-ph",
    reportNumber = "TUM-HEP-1348/21",
    doi = "10.1007/JHEP06(2022)023",
    journal = "JHEP",
    volume = "06",
    pages = "023",
    year = "2022"
}

@article{Easther:2009ft,
    author = "Easther, Richard and Giblin, Jr, John T. and Hui, Lam and Lim, Eugene A.",
    title = "{A New Mechanism for Bubble Nucleation: Classical Transitions}",
    eprint = "0907.3234",
    archivePrefix = "arXiv",
    primaryClass = "hep-th",
    reportNumber = "PI-COSMO-140",
    doi = "10.1103/PhysRevD.80.123519",
    journal = "Phys. Rev. D",
    volume = "80",
    pages = "123519",
    year = "2009"
}

@article{Blasi:2024mtc,
    author = "Blasi, Simone and Mariotti, Alberto",
    title = "{QCD axion strings or seeds?}",
    eprint = "2405.08060",
    archivePrefix = "arXiv",
    primaryClass = "hep-ph",
    doi = "10.21468/SciPostPhys.18.1.016",
    journal = "SciPost Phys.",
    volume = "18",
    number = "1",
    pages = "016",
    year = "2025"
}

@article{Membrado:1989ke,
    author = "Membrado, M. and Abad, J. and Pacheco, A. F. and Sanudo, J.",
    title = "{NEWTONIAN BOSON SPHERES}",
    doi = "10.1103/PhysRevD.40.2736",
    journal = "Phys. Rev. D",
    volume = "40",
    pages = "2736--2738",
    year = "1989"
}

@article{Kaup:1968zz,
    author = "Kaup, David J.",
    title = "{Klein-Gordon Geon}",
    doi = "10.1103/PhysRev.172.1331",
    journal = "Phys. Rev.",
    volume = "172",
    pages = "1331--1342",
    year = "1968"
}

@article{Davidson:2014hfa,
    author = "Davidson, Sacha",
    title = "{Axions: Bose Einstein Condensate or Classical Field?}",
    eprint = "1405.1139",
    archivePrefix = "arXiv",
    primaryClass = "hep-ph",
    doi = "10.1016/j.astropartphys.2014.12.007",
    journal = "Astropart. Phys.",
    volume = "65",
    pages = "101--107",
    year = "2015"
}

@article{Davidson:2013aba,
    author = "Davidson, Sacha and Elmer, Martin",
    title = "{Bose Einstein condensation of the classical axion field in cosmology?}",
    eprint = "1307.8024",
    archivePrefix = "arXiv",
    primaryClass = "hep-ph",
    doi = "10.1088/1475-7516/2013/12/034",
    journal = "JCAP",
    volume = "12",
    pages = "034",
    year = "2013"
}

@article{Guth:2014hsa,
    author = "Guth, Alan H. and Hertzberg, Mark P. and Prescod-Weinstein, C.",
    title = "{Do Dark Matter Axions Form a Condensate with Long-Range Correlation?}",
    eprint = "1412.5930",
    archivePrefix = "arXiv",
    primaryClass = "astro-ph.CO",
    reportNumber = "MIT-CTP-4625",
    doi = "10.1103/PhysRevD.92.103513",
    journal = "Phys. Rev. D",
    volume = "92",
    number = "10",
    pages = "103513",
    year = "2015"
}

@article{Chavanis:2011zi,
    author = "Chavanis, Pierre-Henri",
    title = "{Mass-radius relation of Newtonian self-gravitating Bose-Einstein condensates with short-range interactions: I. Analytical results}",
    eprint = "1103.2050",
    archivePrefix = "arXiv",
    primaryClass = "astro-ph.CO",
    doi = "10.1103/PhysRevD.84.043531",
    journal = "Phys. Rev. D",
    volume = "84",
    pages = "043531",
    year = "2011"
}

@article{Chavanis:2011zm,
    author = "Chavanis, P. H. and Delfini, L.",
    title = "{Mass-radius relation of Newtonian self-gravitating Bose-Einstein condensates with short-range interactions: II. Numerical results}",
    eprint = "1103.2054",
    archivePrefix = "arXiv",
    primaryClass = "astro-ph.CO",
    doi = "10.1103/PhysRevD.84.043532",
    journal = "Phys. Rev. D",
    volume = "84",
    pages = "043532",
    year = "2011"
}

@article{Khlebnikov:1999pt,
    author = "Khlebnikov, S.",
    title = "{Short scale gravitational instability in a disordered Bose gas}",
    eprint = "astro-ph/9911218",
    archivePrefix = "arXiv",
    reportNumber = "PURD-TH-99-08",
    doi = "10.1103/PhysRevD.62.043519",
    journal = "Phys. Rev. D",
    volume = "62",
    pages = "043519",
    year = "2000"
}

@article{Levkov:2018kau,
    author = "Levkov, D. G. and Panin, A. G. and Tkachev, I. I.",
    title = "{Gravitational Bose-Einstein condensation in the kinetic regime}",
    eprint = "1804.05857",
    archivePrefix = "arXiv",
    primaryClass = "astro-ph.CO",
    reportNumber = "INR-TH-2018-005",
    doi = "10.1103/PhysRevLett.121.151301",
    journal = "Phys. Rev. Lett.",
    volume = "121",
    number = "15",
    pages = "151301",
    year = "2018"
}

@article{Erken:2011vv,
    author = "Erken, O. and Sikivie, P. and Tam, H. and Yang, Q.",
    title = "{Axion Dark Matter and Cosmological Parameters}",
    eprint = "1104.4507",
    archivePrefix = "arXiv",
    primaryClass = "astro-ph.CO",
    doi = "10.1103/PhysRevLett.108.061304",
    journal = "Phys. Rev. Lett.",
    volume = "108",
    pages = "061304",
    year = "2012"
}

@article{Sikivie:2009qn,
    author = "Sikivie, P. and Yang, Q.",
    title = "{Bose-Einstein Condensation of Dark Matter Axions}",
    eprint = "0901.1106",
    archivePrefix = "arXiv",
    primaryClass = "hep-ph",
    reportNumber = "UFIFT-HEP-09-1",
    doi = "10.1103/PhysRevLett.103.111301",
    journal = "Phys. Rev. Lett.",
    volume = "103",
    pages = "111301",
    year = "2009"
}

@article{NANOGrav:2023hvm,
    author = "Afzal, Adeela and others",
    collaboration = "NANOGrav",
    title = "{The NANOGrav 15 yr Data Set: Search for Signals from New Physics}",
    eprint = "2306.16219",
    archivePrefix = "arXiv",
    primaryClass = "astro-ph.HE",
    reportNumber = "FERMILAB-PUB-23-589-T",
    doi = "10.3847/2041-8213/acdc91",
    journal = "Astrophys. J. Lett.",
    volume = "951",
    number = "1",
    pages = "L11",
    year = "2023",
    note = "[Erratum: Astrophys.J.Lett. 971, L27 (2024), Erratum: Astrophys.J. 971, L27 (2024)]"
}

@article{Barausse:2020rsu,
    author = "Barausse, Enrico and others",
    title = "{Prospects for Fundamental Physics with LISA}",
    eprint = "2001.09793",
    archivePrefix = "arXiv",
    primaryClass = "gr-qc",
    doi = "10.1007/s10714-020-02691-1",
    journal = "Gen. Rel. Grav.",
    volume = "52",
    number = "8",
    pages = "81",
    year = "2020"
}

@article{Visinelli:2017ooc,
    author = "Visinelli, Luca and Baum, Sebastian and Redondo, Javier and Freese, Katherine and Wilczek, Frank",
    title = "{Dilute and dense axion stars}",
    eprint = "1710.08910",
    archivePrefix = "arXiv",
    primaryClass = "astro-ph.CO",
    reportNumber = "MCTP-17-20A, MIT-CTP-4949, NORDITA-2017-112",
    doi = "10.1016/j.physletb.2017.12.010",
    journal = "Phys. Lett. B",
    volume = "777",
    pages = "64--72",
    year = "2018"
}

@article{Schiappacasse:2017ham,
    author = "Schiappacasse, Enrico D. and Hertzberg, Mark P.",
    title = "{Analysis of Dark Matter Axion Clumps with Spherical Symmetry}",
    eprint = "1710.04729",
    archivePrefix = "arXiv",
    primaryClass = "hep-ph",
    doi = "10.1088/1475-7516/2018/01/037",
    journal = "JCAP",
    volume = "01",
    pages = "037",
    year = "2018",
    note = "[Erratum: JCAP 03, E01 (2018)]"
}

@article{Zakharov-Kuznetsov,
author = {Zakharov, V. and Kuznetsov, E.A.},
year = {2012},
month = {06},
pages = {535},
title = {Solitons and collapses: Two evolution scenarios of nonlinear wave systems},
volume = {55},
journal = {Physics-Uspekhi},
doi = {10.3367/UFNe.0182.201206a.0569}
}

@article{Athron:2023xlk,
    author = "Athron, Peter and Bal{\'a}zs, Csaba and Fowlie, Andrew and Morris, Lachlan and Wu, Lei",
    title = "{Cosmological phase transitions: From perturbative particle physics to gravitational waves}",
    eprint = "2305.02357",
    archivePrefix = "arXiv",
    primaryClass = "hep-ph",
    doi = "10.1016/j.ppnp.2023.104094",
    journal = "Prog. Part. Nucl. Phys.",
    volume = "135",
    pages = "104094",
    year = "2024"
}

@article{Chung-Jukko:2024hod,
    author = "Chung-Jukko, Liina and Lim, Eugene A. and Marsh, David J. E.",
    title = "{Multimessenger signals from compact axion star mergers}",
    eprint = "2403.03774",
    archivePrefix = "arXiv",
    primaryClass = "astro-ph.CO",
    reportNumber = "KCL-PH-TH/2024-14",
    doi = "10.1103/PhysRevD.110.063506",
    journal = "Phys. Rev. D",
    volume = "110",
    number = "6",
    pages = "063506",
    year = "2024"
}

@article{Blum:2024hcs,
    author = "Blum, Kfir and Mirbabayi, Mehrdad",
    title = "{A single-bubble source for gravitational waves in a cosmological phase transition}",
    eprint = "2403.20164",
    archivePrefix = "arXiv",
    primaryClass = "gr-qc",
    doi = "10.1088/1475-7516/2024/08/039",
    journal = "JCAP",
    volume = "08",
    pages = "039",
    year = "2024"
}

@article{Giblin:2010bd,
    author = "Giblin, Jr, John T. and Hui, Lam and Lim, Eugene A. and Yang, I-Sheng",
    title = "{How to Run Through Walls: Dynamics of Bubble and Soliton Collisions}",
    eprint = "1005.3493",
    archivePrefix = "arXiv",
    primaryClass = "hep-th",
    reportNumber = "PI-COSMO-184",
    doi = "10.1103/PhysRevD.82.045019",
    journal = "Phys. Rev. D",
    volume = "82",
    pages = "045019",
    year = "2010"
}

@article{Kobayashi:2018nzh,
    author = "Kobayashi, Takeshi and Ferreira, Pedro G.",
    title = "{Emergent Dark Energy from Dark Matter}",
    eprint = "1801.09658",
    archivePrefix = "arXiv",
    primaryClass = "astro-ph.CO",
    reportNumber = "SISSA-02-2018-FISI",
    doi = "10.1103/PhysRevD.97.121301",
    journal = "Phys. Rev. D",
    volume = "97",
    number = "12",
    pages = "121301",
    year = "2018"
}

@article{Agrawal:2023cgp,
    author = "Agrawal, Prateek and Blasi, Simone and Mariotti, Alberto and Nee, Michael",
    title = "{Electroweak phase transition with a double well done doubly well}",
    eprint = "2312.06749",
    archivePrefix = "arXiv",
    primaryClass = "hep-ph",
    reportNumber = "DESY-23-208",
    doi = "10.1007/JHEP06(2024)089",
    journal = "JHEP",
    volume = "06",
    pages = "089",
    year = "2024"
}

@article{Blasi:2023rqi,
    author = "Blasi, Simone and Jinno, Ryusuke and Konstandin, Thomas and Rubira, Henrique and Stomberg, Isak",
    title = "{Gravitational waves from defect-driven phase transitions: domain walls}",
    eprint = "2302.06952",
    archivePrefix = "arXiv",
    primaryClass = "astro-ph.CO",
    doi = "10.1088/1475-7516/2023/10/051",
    journal = "JCAP",
    volume = "10",
    pages = "051",
    year = "2023"
}

@article{DiLuzio:2021gos,
    author = "Di Luzio, Luca and Gavela, Belen and Quilez, Pablo and Ringwald, Andreas",
    title = "{Dark matter from an even lighter QCD axion: trapped misalignment}",
    eprint = "2102.01082",
    archivePrefix = "arXiv",
    primaryClass = "hep-ph",
    reportNumber = "DESY 21-011, DESY-21-011, IFT-UAM/CSIC-20-144, FTUAM-20-21",
    doi = "10.1088/1475-7516/2021/10/001",
    journal = "JCAP",
    volume = "10",
    pages = "001",
    year = "2021"
}

@article{Higaki:2016yqk,
    author = "Higaki, Tetsutaro and Jeong, Kwang Sik and Kitajima, Naoya and Takahashi, Fuminobu",
    title = "{Quality of the Peccei-Quinn symmetry in the Aligned QCD Axion and Cosmological Implications}",
    eprint = "1603.02090",
    archivePrefix = "arXiv",
    primaryClass = "hep-ph",
    reportNumber = "TU-1017, IPMU16-0030, APCTP-PRE-2016-006, PNUTP-16-A11",
    doi = "10.1007/JHEP06(2016)150",
    journal = "JHEP",
    volume = "06",
    pages = "150",
    year = "2016"
}

@article{Shkerin:2021zbf,
    author = "Shkerin, Andrey and Sibiryakov, Sergey",
    title = "{Black hole induced false vacuum decay from first principles}",
    eprint = "2105.09331",
    archivePrefix = "arXiv",
    primaryClass = "hep-th",
    reportNumber = "FTPI-MINN-21-07, UMN-TH-4014/21, INR-TH-2021-011",
    doi = "10.1007/JHEP11(2021)197",
    journal = "JHEP",
    volume = "11",
    pages = "197",
    year = "2021"
}

@article{Coleman:1977py,
      author         = "Coleman, Sidney R.",
      title          = "{The Fate of the False Vacuum. 1. Semiclassical Theory}",
      journal        = "Phys. Rev.",
      volume         = "D15",
      year           = "1977",
      pages          = "2929-2936",
      doi            = "10.1103/PhysRevD.15.2929, 10.1103/PhysRevD.16.1248",
      note           = "[Erratum: Phys. Rev.D16,1248(1977)]",
      reportNumber   = "HUTP-77-A004",
      SLACcitation   = "%%CITATION = PHRVA,D15,2929;%%"
}

@article{Blasi:2022woz,
    author = "Blasi, Simone and Mariotti, Alberto",
    title = "{Domain walls seeding the electroweak phase transition}",
    eprint = "2203.16450",
    archivePrefix = "arXiv",
    primaryClass = "hep-ph",
    month = "3",
    year = "2022"
}

@article{LINDE1980289,
title = {Infrared problem in the thermodynamics of the Yang-Mills gas},
journal = {Physics Letters B},
volume = {96},
number = {3},
pages = {289-292},
year = {1980},
issn = {0370-2693},
doi = {https://doi.org/10.1016/0370-2693(80)90769-8},
url = {https://www.sciencedirect.com/science/article/pii/0370269380907698},
author = {A.D. Linde}
}

@article{Wainwright:2011kj,
      author         = "Wainwright, Carroll L.",
      title          = "{CosmoTransitions: Computing Cosmological Phase
                        Transition Temperatures and Bubble Profiles with Multiple
                        Fields}",
      journal        = "Comput. Phys. Commun.",
      volume         = "183",
      year           = "2012",
      pages          = "2006-2013",
      doi            = "10.1016/j.cpc.2012.04.004",
      eprint         = "1109.4189",
      archivePrefix  = "arXiv",
      primaryClass   = "hep-ph",
      SLACcitation   = "%%CITATION = ARXIV:1109.4189;%%"
}

@article{Linde:1980tt,
      author         = "Linde, Andrei D.",
      title          = "{Fate of the False Vacuum at Finite Temperature: Theory
                        and Applications}",
      journal        = "Phys. Lett.",
      volume         = "100B",
      year           = "1981",
      pages          = "37-40",
      doi            = "10.1016/0370-2693(81)90281-1",
      reportNumber   = "LEBEDEV-80-92",
      SLACcitation   = "%%CITATION = PHLTA,100B,37;%%"
}

@article{Marsh:2015xka,
    author = "Marsh, David J. E.",
    title = "{Axion Cosmology}",
    eprint = "1510.07633",
    archivePrefix = "arXiv",
    primaryClass = "astro-ph.CO",
    reportNumber = "KCL-PH-TH-2015-50",
    doi = "10.1016/j.physrep.2016.06.005",
    journal = "Phys. Rept.",
    volume = "643",
    pages = "1--79",
    year = "2016"
}

@article{Guada:2020xnz,
    author = "Guada, Victor and Nemev\v{s}ek, Miha and Pintar, Matev\v{z}",
    title = "{FindBounce: Package for multi-field bounce actions}",
    eprint = "2002.00881",
    archivePrefix = "arXiv",
    primaryClass = "hep-ph",
    doi = "10.1016/j.cpc.2020.107480",
    journal = "Comput. Phys. Commun.",
    volume = "256",
    pages = "107480",
    year = "2020"
}

@book{Coleman:1985rnk,
    author = "Coleman, Sidney",
    title = "{Aspects of Symmetry}: {Selected Erice Lectures}",
    doi = "10.1017/CBO9780511565045",
    isbn = "978-0-521-31827-3",
    publisher = "Cambridge University Press",
    address = "Cambridge, U.K.",
    year = "1985"
}

@article{Ralegankar:2024zjd,
    author = "Ralegankar, Pranjal and Perri, Daniele and Kobayashi, Takeshi",
    title = "{Gravothermalizing into primordial black holes, boson stars, and cannibal stars}",
    eprint = "2410.18948",
    archivePrefix = "arXiv",
    primaryClass = "astro-ph.CO",
    doi = "10.1103/xpwl-w5zk",
    journal = "Phys. Rev. D",
    volume = "112",
    number = "8",
    pages = "083019",
    year = "2025"
}

@article{Gorghetto:2024vnp,
    author = "Gorghetto, Marco and Hardy, Edward and Villadoro, Giovanni",
    title = "{More axion stars from strings}",
    eprint = "2405.19389",
    archivePrefix = "arXiv",
    primaryClass = "hep-ph",
    reportNumber = "DESY-24-075",
    doi = "10.1007/JHEP08(2024)126",
    journal = "JHEP",
    volume = "08",
    pages = "126",
    year = "2024"
}

@article{Dmitriev:2023ipv,
    author = "Dmitriev, A. S. and Levkov, D. G. and Panin, A. G. and Tkachev, I. I.",
    title = "{Self-Similar Growth of Bose Stars}",
    eprint = "2305.01005",
    archivePrefix = "arXiv",
    primaryClass = "astro-ph.CO",
    reportNumber = "INR-TH-2023-006",
    doi = "10.1103/PhysRevLett.132.091001",
    journal = "Phys. Rev. Lett.",
    volume = "132",
    number = "9",
    pages = "091001",
    year = "2024"
}

@article{Chan:2022bkz,
    author = "Chan, James Hung-Hsu and Sibiryakov, Sergey and Xue, Wei",
    title = "{Condensation and evaporation of boson stars}",
    eprint = "2207.04057",
    archivePrefix = "arXiv",
    primaryClass = "astro-ph.CO",
    doi = "10.1007/JHEP01(2024)071",
    journal = "JHEP",
    volume = "01",
    pages = "071",
    year = "2024"
}

@article{Kolb:1993zz,
    author = "Kolb, Edward W. and Tkachev, Igor I.",
    title = "{Axion miniclusters and Bose stars}",
    eprint = "hep-ph/9303313",
    archivePrefix = "arXiv",
    reportNumber = "FERMILAB-PUB-93-066-A",
    doi = "10.1103/PhysRevLett.71.3051",
    journal = "Phys. Rev. Lett.",
    volume = "71",
    pages = "3051--3054",
    year = "1993"
}

@article{Eggemeier:2019jsu,
    author = "Eggemeier, Benedikt and Niemeyer, Jens C.",
    title = "{Formation and mass growth of axion stars in axion miniclusters}",
    eprint = "1906.01348",
    archivePrefix = "arXiv",
    primaryClass = "astro-ph.CO",
    doi = "10.1103/PhysRevD.100.063528",
    journal = "Phys. Rev. D",
    volume = "100",
    number = "6",
    pages = "063528",
    year = "2019"
}

@article{Chen:2021oot,
    author = "Chen, Jiajun and Du, Xiaolong and Lentz, Erik W. and Marsh, David J. E.",
    title = "{Relaxation times for Bose-Einstein condensation by self-interaction and gravity}",
    eprint = "2109.11474",
    archivePrefix = "arXiv",
    primaryClass = "astro-ph.CO",
    doi = "10.1103/PhysRevD.106.023009",
    journal = "Phys. Rev. D",
    volume = "106",
    number = "2",
    pages = "023009",
    year = "2022"
}

@article{Flores:2020drq,
    author = "Flores, Marcos M. and Kusenko, Alexander",
    title = "{Primordial Black Holes from Long-Range Scalar Forces and Scalar Radiative Cooling}",
    eprint = "2008.12456",
    archivePrefix = "arXiv",
    primaryClass = "astro-ph.CO",
    reportNumber = "IPMU20-0092",
    doi = "10.1103/PhysRevLett.126.041101",
    journal = "Phys. Rev. Lett.",
    volume = "126",
    number = "4",
    pages = "041101",
    year = "2021"
}

@article{Agrawal:2022hnf,
    author = "Agrawal, Prateek and Nee, Michael",
    title = "{The Boring Monopole}",
    eprint = "2202.11102",
    archivePrefix = "arXiv",
    primaryClass = "hep-ph",
    doi = "10.21468/SciPostPhys.13.3.049",
    journal = "SciPost Phys.",
    volume = "13",
    number = "3",
    pages = "049",
    year = "2022"
}

@article{Eby:2015hsq,
    author = "Eby, Joshua and Kouvaris, Chris and Nielsen, Niklas Gr{\o}nlund and Wijewardhana, L. C. R.",
    title = "{Boson Stars from Self-Interacting Dark Matter}",
    eprint = "1511.04474",
    archivePrefix = "arXiv",
    primaryClass = "hep-ph",
    doi = "10.1007/JHEP02(2016)028",
    journal = "JHEP",
    volume = "02",
    pages = "028",
    year = "2016"
}

@article{Liddle:1992fmk,
    author = "Liddle, Andrew R. and Madsen, Mark S.",
    title = "{The Structure and formation of boson stars}",
    reportNumber = "SUSSEX-AST-92-2-1",
    doi = "10.1142/S0218271892000057",
    journal = "Int. J. Mod. Phys. D",
    volume = "1",
    pages = "101--144",
    year = "1992"
}

@article{Colpi:1986ye,
    author = "Colpi, M. and Shapiro, S. L. and Wasserman, I.",
    title = "{Boson Stars: Gravitational Equilibria of Selfinteracting Scalar Fields}",
    doi = "10.1103/PhysRevLett.57.2485",
    journal = "Phys. Rev. Lett.",
    volume = "57",
    pages = "2485--2488",
    year = "1986"
}

@article{Zhang:2018slz,
    author = "Zhang, Hong",
    title = "{Axion Stars}",
    eprint = "1810.11473",
    archivePrefix = "arXiv",
    primaryClass = "hep-ph",
    doi = "10.3390/sym12010025",
    journal = "Symmetry",
    volume = "12",
    number = "1",
    pages = "25",
    year = "2019"
}

@article{Yajnik:1986tg,
    author = "Yajnik, U. A.",
    title = "{PHASE TRANSITION INDUCED BY COSMIC STRINGS}",
    doi = "10.1103/PhysRevD.34.1237",
    journal = "Phys. Rev. D",
    volume = "34",
    pages = "1237--1240",
    year = "1986"
}

@article{Steinhardt:1981ec,
    author = "Steinhardt, Paul Joseph",
    title = "{Monopole and Vortex Dissociation and Decay of the False Vacuum}",
    reportNumber = "HUTP-80/A088",
    doi = "10.1016/0550-3213(81)90449-1",
    journal = "Nucl. Phys. B",
    volume = "190",
    pages = "583--616",
    year = "1981"
}

@article{Kumar:2010mv,
    author = "Kumar, Brijesh and Paranjape, M. B. and Yajnik, U. A.",
    title = "{Fate of the false monopoles: Induced vacuum decay}",
    eprint = "1006.0693",
    archivePrefix = "arXiv",
    primaryClass = "hep-th",
    reportNumber = "UDEM-GPP-TH-10-189",
    doi = "10.1103/PhysRevD.82.025022",
    journal = "Phys. Rev. D",
    volume = "82",
    pages = "025022",
    year = "2010"
}

@article{Hosotani:1982ii,
    author = "Hosotani, Yutaka",
    title = "{Impurities in the Early Universe}",
    reportNumber = "UPR-0205T",
    doi = "10.1103/PhysRevD.27.789",
    journal = "Phys. Rev. D",
    volume = "27",
    pages = "789",
    year = "1983"
}

@article{Preskill:1992ck,
    author = "Preskill, John and Vilenkin, Alexander",
    title = "{Decay of metastable topological defects}",
    eprint = "hep-ph/9209210",
    archivePrefix = "arXiv",
    reportNumber = "HUTP-92-A018, CALT-68-1786",
    doi = "10.1103/PhysRevD.47.2324",
    journal = "Phys. Rev. D",
    volume = "47",
    pages = "2324--2342",
    year = "1993"
}

@article{Graham:2015cka,
    author = "Graham, Peter W. and Kaplan, David E. and Rajendran, Surjeet",
    title = "{Cosmological Relaxation of the Electroweak Scale}",
    eprint = "1504.07551",
    archivePrefix = "arXiv",
    primaryClass = "hep-ph",
    doi = "10.1103/PhysRevLett.115.221801",
    journal = "Phys. Rev. Lett.",
    volume = "115",
    number = "22",
    pages = "221801",
    year = "2015"
}

@article{Choi:2015fiu,
    author = "Choi, Kiwoon and Im, Sang Hui",
    title = "{Realizing the relaxion from multiple axions and its UV completion with high scale supersymmetry}",
    eprint = "1511.00132",
    archivePrefix = "arXiv",
    primaryClass = "hep-ph",
    reportNumber = "CTPU-15-16",
    doi = "10.1007/JHEP01(2016)149",
    journal = "JHEP",
    volume = "01",
    pages = "149",
    year = "2016"
}

@article{Kaplan:2015fuy,
    author = "Kaplan, David E. and Rattazzi, Riccardo",
    title = "{Large field excursions and approximate discrete symmetries from a clockwork axion}",
    eprint = "1511.01827",
    archivePrefix = "arXiv",
    primaryClass = "hep-ph",
    doi = "10.1103/PhysRevD.93.085007",
    journal = "Phys. Rev. D",
    volume = "93",
    number = "8",
    pages = "085007",
    year = "2016"
}

@article{Croon:2018ftb,
    author = "Croon, Djuna and Gleiser, Marcelo and Mohapatra, Sonali and Sun, Chen",
    title = "{Gravitational Radiation Background from Boson Star Binaries}",
    eprint = "1802.08259",
    archivePrefix = "arXiv",
    primaryClass = "hep-ph",
    doi = "10.1016/j.physletb.2018.03.055",
    journal = "Phys. Lett. B",
    volume = "783",
    pages = "158--162",
    year = "2018"
}

@article{Hiscock:1987hn,
    author = "Hiscock, W. A.",
    title = "{CAN BLACK HOLES NUCLEATE VACUUM PHASE TRANSITIONS?}",
    doi = "10.1103/PhysRevD.35.1161",
    journal = "Phys. Rev. D",
    volume = "35",
    pages = "1161--1170",
    year = "1987"
}

@article{Gregory:2013hja,
    author = "Gregory, Ruth and Moss, Ian G. and Withers, Benjamin",
    title = "{Black holes as bubble nucleation sites}",
    eprint = "1401.0017",
    archivePrefix = "arXiv",
    primaryClass = "hep-th",
    reportNumber = "DCPT-13-43",
    doi = "10.1007/JHEP03(2014)081",
    journal = "JHEP",
    volume = "03",
    pages = "081",
    year = "2014"
}

@article{Mukaida:2017bgd,
    author = "Mukaida, Kyohei and Yamada, Masaki",
    title = "{False Vacuum Decay Catalyzed by Black Holes}",
    eprint = "1706.04523",
    archivePrefix = "arXiv",
    primaryClass = "hep-th",
    reportNumber = "IPMU-17-0069",
    doi = "10.1103/PhysRevD.96.103514",
    journal = "Phys. Rev. D",
    volume = "96",
    number = "10",
    pages = "103514",
    year = "2017"
}

\end{document}